\documentclass[10pt,journal,compsoc]{IEEEtran}
\ifCLASSOPTIONcompsoc
\else
  \usepackage[noadjust]{cite}
\fi
\ifCLASSINFOpdf
  \usepackage[pdftex]{graphicx}
\else
\fi

\usepackage{algorithm,algorithmic}

\usepackage[none]{hyphenat}

\begin{document}
%
% paper title
% Titles are generally capitalized except for words such as a, an, and, as,
% at, but, by, for, in, nor, of, on, or, the, to and up, which are usually
% not capitalized unless they are the first or last word of the title.
% Linebreaks \\ can be used within to get better formatting as desired.
% Do not put math or special symbols in the title.
\title{An Event-Driven Approach To Genotype Imputation On A Custom RISC-V FPGA Cluster}
%
%
% author names and IEEE memberships
% note positions of commas and nonbreaking spaces ( ~ ) LaTeX will not break
% a structure at a ~ so this keeps an author's name from being broken across
% two lines.
% use \thanks{} to gain access to the first footnote area
% a separate \thanks must be used for each paragraph as LaTeX2e's \thanks
% was not built to handle multiple paragraphs
%
%
%\IEEEcompsocitemizethanks is a special \thanks that produces the bulleted
% lists the Computer Society journals use for "first footnote" author
% affiliations. Use \IEEEcompsocthanksitem which works much like \item
% for each affiliation group. When not in compsoc mode,
% \IEEEcompsocitemizethanks becomes like \thanks and
% \IEEEcompsocthanksitem becomes a line break with idention. This
% facilitates dual compilation, although admittedly the differences in the
% desired content of \author between the different types of papers makes a
% one-size-fits-all approach a daunting prospect. For instance, compsoc 
% journal papers have the author affiliations above the "Manuscript
% received ..."  text while in non-compsoc journals this is reversed. Sigh.

\author{Jordan~Morris, Ashur Rafiev, Graeme M. Bragg, Mark L. Vousden, 
        David B. Thomas, Alex~Yakovlev,~\IEEEmembership{Fellow,~IEEE,}
        and~Andrew~D.~Brown,~\IEEEmembership{Senior~Member,~IEEE}% <-this % stops a space
\IEEEcompsocitemizethanks{\IEEEcompsocthanksitem Jordan Morris, Ashur Rafiev and Alex Yakovlev are with the $\mu$Systems design group, Newcastle University, Newcastle, NE1 7RU .\protect\\
% note need leading \protect in front of \\ to get a newline within \thanks as
% \\ is fragile and will error, could use \hfil\break instead.
E-mail: \{jordan.morris, ashur.rafiev, alex.yakovlev\}@newcastle.ac.uk
\IEEEcompsocthanksitem Graeme M. Bragg, Mark L. Vousden and A. D. Brown are with the Department of Electronics and Computer Science, University of Southampton, SO17 1BJ, UK
\protect\\ 
E-mail: \{gmb, m.vousden, adb@ecs.soton.ac.uk\}
\IEEEcompsocthanksitem David B. Thomas is with the Electronic Engineering Department at Imperial College London, South Kensington Campus, London SW7
2AZ, UK
\protect\\ 
E-mail: \{d.thomas1@imperial.ac.uk\}

}% <-this % stops a space
}
\IEEEtitleabstractindextext{%
\begin{abstract}
This paper proposes an event-driven solution to genotype imputation, a technique used to statistically infer missing genetic markers in DNA. The work implements the widely accepted Li and Stephens model, primary contributor to the computational complexity of modern x86 solutions, in an attempt to determine whether further investigation of the application is warranted in the event-driven domain. The model is implemented using graph-based Hidden Markov Modeling and executed as a customized forward/backward dynamic programming algorithm. The solution uses an event-driven paradigm to map the algorithm to thousands of concurrent cores, where events are small messages that carry both control and data within the algorithm. The design of a single processing element is discussed. This is then extended across multiple FPGAs and executed on a custom RISC-V NoC FPGA cluster called POETS. Results demonstrate how the algorithm scales over increasing hardware resources and a 48 FPGA run demonstrates a 270X reduction in wall-clock processing time when compared to a single-threaded x86 solution. Optimisation of the algorithm via linear interpolation is then introduced and tested, with results demonstrating a wall-clock reduction time of $\mathtt{\sim}$5 orders of magnitude when compared to a similarly optimised x86 solution. 
\end{abstract}

% Note that keywords are not normally used for peerreview papers.
}

% make the title area
\maketitle

% To allow for easy dual compilation without having to reenter the
% abstract/keywords data, the \IEEEtitleabstractindextext text will
% not be used in maketitle, but will appear (i.e., to be "transported")
% here as \IEEEdisplaynontitleabstractindextext when compsoc mode
% is not selected <OR> if conference mode is selected - because compsoc
% conference papers position the abstract like regular (non-compsoc)
% papers do!
\IEEEdisplaynontitleabstractindextext
% \IEEEdisplaynontitleabstractindextext has no effect when using
% compsoc under a non-conference mode.

% For peer review papers, you can put extra information on the cover
% page as needed:
% \ifCLASSOPTIONpeerreview
% \begin{center} \bfseries EDICS Category: 3-BBND \end{center}
% \fi
%
% For peerreview papers, this IEEEtran command inserts a page break and
% creates the second title. It will be ignored for other modes.
\IEEEpeerreviewmaketitle

\ifCLASSOPTIONcompsoc
\IEEEraisesectionheading{\section{Introduction}\label{sec:introduction}}
\else
\section{Introduction}
\label{sec:introduction}
\fi
% Computer Society journal (but not conference!) papers do something unusual
% with the very first section heading (almost always called "Introduction").
% They place it ABOVE the main text! IEEEtran.cls does not automatically do
% this for you, but you can achieve this effect with the provided
% \IEEEraisesectionheading{} command. Note the need to keep any \label that
% is to refer to the section immediately after \section in the above as
% \IEEEraisesectionheading puts \section within a raised box.

% The very first letter is a 2 line initial drop letter followed
% by the rest of the first word in caps (small caps for compsoc).
% 
% form to use if the first word consists of a single letter:
% \IEEEPARstart{A}{demo} file is ....
% 
% form to use if you need the single drop letter followed by
% normal text (unknown if ever used by the IEEE):
% \IEEEPARstart{A}{}demo file is ....
% 
% Some journals put the first two words in caps:
% \IEEEPARstart{T}{his demo} file is ....
% 
% Here we have the typical use of a "T" for an initial drop letter
% and "HIS" in caps to complete the first word.
\IEEEPARstart{O}{ne} of the most important concepts in modern human genetics is the ability to link observable traits back to their causative genes and identify associated allele variations. This process is often the first step in understanding the origin of genetically-linked medical conditions and frequently leads onto therapeutic design for heritable human diseases\cite{MAHJANI}. \\
\indent One methodology to locate causative genetic loci that has gained in popularity and accuracy over the past decade is called a Genome Wide Association Study (GWAS). These studies sample marker locations spread across the entire genome and are conducted on populations at large (1k-100k participants). By separating the sampled data based on an observable trait, differences in the allele frequencies of the partitioned groups can be used to identify loci with statistically significant correlation to the trait under investigation \cite{MACARTHUR}. \\
\indent The accuracy of GWAS improves as the number of participants increases, due to the increase in the probability that the data includes rare genetic variants. This has fuelled an exponential trend. HapMap3\cite{Altshuler} (2010) used around 1K haplotypes, yet sample sizes of 10M haplotypes are predicted within a decade\cite{HUANG, ABECASIS, ABECASIS2, HUANG2, AUTON}. Moreover, genotyping-by-chip (GBC) technology has also improved exponentially over the past decade. HapMap3\cite{Altshuler} sampled $\mathtt{\sim}$1.4M marker loci, whereas more recent studies such as TopMED\cite{TOPMED} sampled $\mathtt{\sim}$240M. Even assuming a bulk purchase GBC price of \$40 per participant, GWAS are expensive endeavours. Moreover, at the current rate of exponential improvement, their results would only be relevant for $\mathtt{\sim}$18 months before being superseded by newer studies with greater participation and better technology. \\
\indent To extend the relevance of the data gathered during GWAS, a model to statistically infer predicted markers from GWAS run on newer technology has been proposed\cite{LI}\cite{SCHEET}. This takes advantage of the probabilistic nature of human genetics and is known as genotype imputation. The problem corresponds to solving a customized Hidden Markov Model (HMM)\cite{RABINER} using a unique permutation of the forward/backward dynamic programming algorithm. The state space is constructed as a 2D reference panel generated using data from a recent GWAS. Each HMM state is labelled with an allele from the reference panel. Once the posterior probabilities of the model have been calculated, they may be summed based on their allele label and the allele with the highest accumulated probability may be inferred to reside at a particular marker location. The overall process may be considered a statistical 'fill in the blanks' exercise, increasing the marker count in older GWAS up to the latest technology, thereby increasing the accuracy and relevance of pre-existing GWAS data.\\ 
\indent Several x86 implementations have been explored in the literature\cite{BROWNING, DAS, HOWIE, LI2}, yet the exponential increase in the numbers of markers and participants has the potential to test the boundaries of modern x86 solutions in the near future. Even moderate reference panel sizes can generate memory requirements upwards of 10's GB per thread. Moreover, the wall-clock run-times for reasonably large reference panels are typically measured in days. These issues require significant development, both architecturally and algorithmically, to be resolved. \\
\indent As high performance computing (HPC) clusters become ever more expensive, modular FPGA-based workflows have gained favor in providing scalable solutions, particularly for algorithms that inherently break down into a large number of discrete, orchestrated processing elements. This work leverages a RISC-V FPGA-based NoC cluster called POETS\cite{BROWN} to evaluate a scalable distributed solution to genotype imputation based on HMM. The contributions of this paper are:-
\begin{itemize}
  \item Mapping of the widely accepted Li and Stephens model\cite{LI} into a new event-driven algorithm.
  \item An implementation of this new algorithm on the POETS cluster, that is concurrent and parallel at both the inter- and intra-FPGA level.
  \item An evaluation of how this new algorithm scales into expanding hardware resources (more FPGAs).
  \item An evaluation of how this new algorithm performs using 48 networked FPGAs, showing up to a 270X speed up over a single-threaded x86 solution.
  \item An evaluation of how linear interpolation can optimise the new algorithm, with 48 networked FPGAs showing up to $\mathtt{\sim}$5 orders speed up over a single-threaded x86 solution.
\end{itemize}

The paper consists of 7 sections. Section 2 provides a brief overview of the technology used to gather the genetic information used in genotype imputation. Section 3 provides a description of how genotype imputation is performed, including construction of the reference panels and the underlying programming model. Section 4 outlines the POETS architecture, providing a description of the physical hardware, the software stack and how the architecture is configured with application graphs. Section 5 introduces the proposed event-driven solution to genotype imputation, including how the application graph is created from the reference panel, a description of the algorithm and a staged walk-through of how information travels and is processed through the graph. Optimisation via linear interpolation is also described. Section 6 describes the experiments conducted to determine the performance of the proposed solution in comparison to an x86 implementation. Results with and without linear interpolation optimisation are presented and explained. Section 7 concludes the paper.

\section{Genotyping-by-chip}

Even with the advent of next generation sequencing, the process of sequencing an entire human genome is still relatively expensive ($\mathtt{\sim}$\$1K with appropriate services to an appropriate depth). Moreover, sequencing your full genome and a randomly chosen neighbours would result in a base pair sequence that is 99.5\% identical, creating an overall information redundancy of the same order. \\ 
\indent A cheaper methodology to sample genetic information was therefore devised. This is called Genotyping-By-Chip (GBC)\cite{PERKEL}. GBC technology uses a silicon die consisting of a microarray of beads. Attached to each bead are the complementary base pairs of around 50 nucleotides leading up to a particular location of interest. This location is called a marker. DNA from a subject is amplified using polymerase chain reaction and then broken into smaller fragments. The beads are then incubated with these fragments and DNA polymerase added such that the DNA strands attach to the beads using complementary base pairing up to the marker location. Modified, fluorescently tagged nucleotides are then added, and the DNA polymerase adds on the next nucleotide corresponding to the marker location. Once this has been completed, a microfine laser light is simply used to read back the fluorescence of the bead and the base type at the marker location in question can be determined (A - Adenine, T - Thymine, C - Cytosine, G - Guanine). \\
\indent By selecting marker locations of interest across the entire genome, the resultant data after pre-phasing\cite{HOWIE2} is two haplotypes with millions of locations corresponding to the two sets of chromosomes each person has. Crucially, this process is an order of magnitude cheaper than sequencing the entire genome, enabling bulk sampling of populations as a whole. This then enables methodologies such as Genome Wide Association Studies (GWAS) to be conducted.

\section{Genotype Imputation}

\subsection{Reference Panel Construction}

\begin{figure}[h!]
\begin{center}
	\includegraphics[scale=0.64]{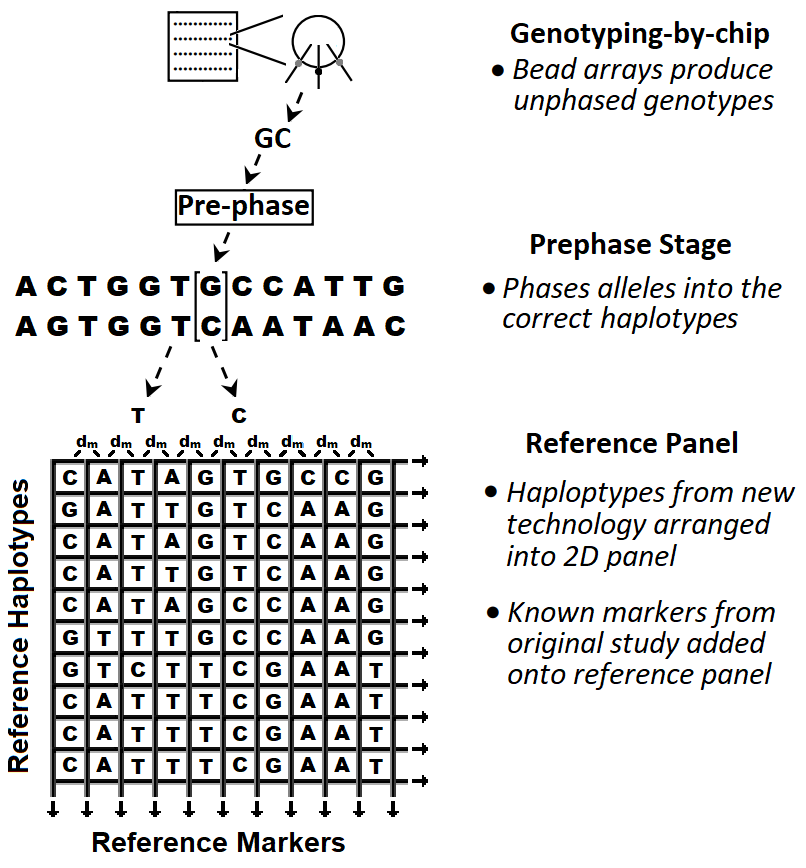}
\end{center}
\caption{Reference Panel Construction}
\label{fig:refpanel}
\end{figure}

\indent The first stage of genotype imputation is constructing the HMM state space. This is constructed as a simple 2D array with haplotypes from the latest GWAS study ‘stacked’ in the vertical dimension, aligned at the sampled marker locations which are labelled in the horizontal dimension. These are referred to as the reference haplotypes and reference markers respectively. Each state in the matrix is labelled with the allele that was determined to be at that marker location for the corresponding reference haplotype. This 2D matrix is called a reference panel. The haplotype from the older data that one is attempting to ‘fill in the blanks’ for is called the target haplotype. Markers from the target haplotype are annotated onto the reference panel at their corresponding marker locations as the process is executed. An overview of this process may be seen in Figure \ref{fig:refpanel} (Note: $d_{m}$ is the genetic distance between adjacent markers and is explained in the next section). \\

\subsection{Imputation Model}
\label{section:model}

\indent The underlying scientific model for genotype imputation was proposed by Li and Stephens\cite{LI} and relies on the concept of linkage disequilibrium. In essence, markers are more likely to be inherited together as the genetic distance between them decreases. This genetic distance is denoted in the model by the term $d_{m}$. It must be noted that although the GBC technology chooses marker loci for an even distribution across the genome, the genetic distances between each pair of marker loci are slightly different. A Tau factor may be derived from the genetic distance as follows:
\begin{equation}
\tau_{m} = 1 - e^{-4N_{e}d_{m}/|H|}
\label{tau}
\end{equation}
whereby $|H|$ is the number of haplotypes and $N_{e}$ represents the effective population size (simply a constant in the model). \\
The model also relies on genetic recombination, where discontinuities in a segment of DNA have resulted from the exchange of DNA segments during meiosis. The above Tau factor may then be used to determine the probability of remaining on the same haplotype, as calculated below:
\begin{equation}
(1-\tau_{m}) + (\tau_{m})/|H|
\label{same}
\end{equation}
or jumping to a different haplotype:
\begin{equation}
(\tau_{m})/|H|
\label{diff}
\end{equation}
\indent The purpose of the algorithm is to determine the probability of being in each state. This requires information from a forward pass and a backward pass through the state space. These are termed as alpha and beta respectively and may by derived as:
\begin{equation}
\alpha_{m+1}(j) = \left[ \sum_{i=1}^{H} \alpha_{m}(i) \alpha_{ij} \right] b_{j}(O_{m+1})\label{alpha}
\end{equation}
\begin{equation}
\beta_{m}(i) = \sum_{j=1}^{H} \alpha_{ij} b_{j}(O_{m+1}) \beta_{m+1}(j)
\label{beta}
\end{equation}
 whereby $\alpha_{m}(i)$ represents an alpha from a given state in the previous marker location (column), $\beta_{m+1}(j)$ represents a beta from a given state in the subsequent marker location (column), $\alpha_{ij}$ is the transition probability of jumping from that state to the current state being calculated ((\ref{same}) or (\ref{diff})) and $b_{j} (O_{m+1})$ is the emission probability of making a particular observation in that state. The emission probability is defined in terms of an error rate $e$ (1/10000 in the model). If the marker loci has an annotated base from the target haplotype, the error rate applied depends on whether the annotated base matches that of the state. If so, the error rate applied is:
\begin{equation}
1 - e
\label{match}
\end{equation}
Otherwise, if the bases mismatch:
\begin{equation}
e
\label{mismatch}
\end{equation}
 If there is no annotated base at that location, the emission probability is assumed to be 1 and the term falls out of the equation. \\
\indent The model is initialized by setting all alpha values in the first marker location (column) to $1/|H|$ and all beta values in the final marker location (column) to 1. The alpha values are calculated from left to right using (\ref{alpha}) and the beta values are calculated from right to left using (\ref{beta}). \\
\indent Once a state has both an alpha and a beta value, these are multiplied together to produce a posterior probability. This represents the probability of being in that particular state at that marker location. The posterior probabilities may then be summed based on their base labels in the reference panel to generate an overall probability of that base occurring at that marker location. 

\section{POETS Architecture}
\subsection{Overview}

\indent POETS (Partially Ordered Event Triggered Systems) is both a development-model and concrete hardware-software stack for developing event-driven solutions, with a message-passing architecture that is subtly different from its peers. POETS was designed from the ground up for large, graph-based simulation problems. It is therefore ideally suited for problems that can be broken down into graphs of thousands or millions of concurrent state-machines (vertices), each connected to neighbouring devices by communication channels (edges) over which state-machines can send and receive events. An event is a small, atomic, asynchronous packet (e.g. 64 bytes) which carries both control and data needed to solve the problem. Through the concurrent and asynchronous exchange of millions or billions of messages, the entire graph of state-machines will calculate the global solution. \\
\indent Graphs can represent a physical topology, such as 2/3-dimensional space, which may be decomposed into a large collection of cells. A device may then be allocated to govern the behaviour in each unit cell. These graphs would inherently form a regular structure. However, no restrictions are placed upon the graph by the POETS infrastructure and therefore it is also capable of running graph-based problems of an irregular or arbitrary nature. Moreover, the behaviour of each device is governed by a set of handlers, and these need not be the same for all devices, allowing for graphs with vertices of a heterogeneous nature. The execution of a handler is solely dependent on the arrival of an event. The handler may then alter the persistent internal state of the device, emit events of its own or simply decide no further action is required. Once this has completed, execution ceases awaiting the arrival of the next event. The entire compute trajectory is therefore event-driven.

\subsection{Compute Stack}
\begin{figure}[h!]
\begin{center}
	\includegraphics[scale=0.69]{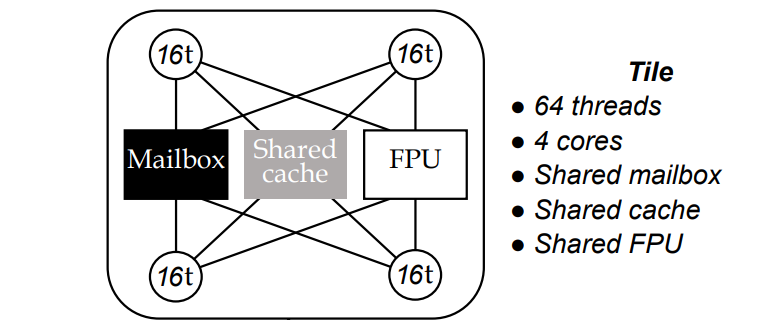}
\end{center}
\caption{Tinsel Tile}
\label{fig:hierarchy1}
\end{figure}

\begin{figure}[h!]
\begin{center}
	\includegraphics[scale=0.69]{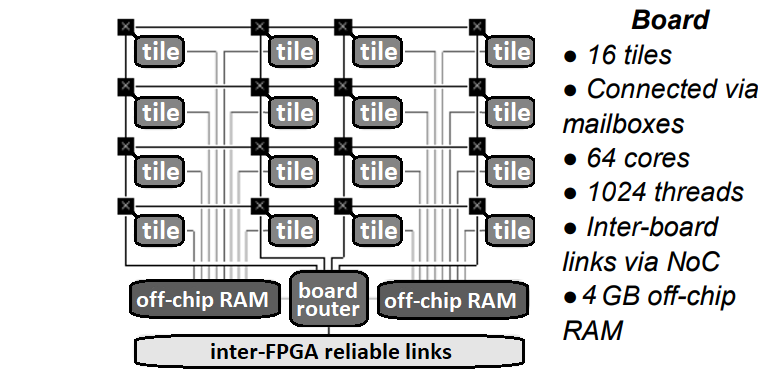}
\end{center}
\caption{Stratix-V FPGA Layout}
\label{fig:hierarchy2}
\end{figure}

\begin{figure}[h!]
\begin{center}
	\includegraphics[scale=0.72]{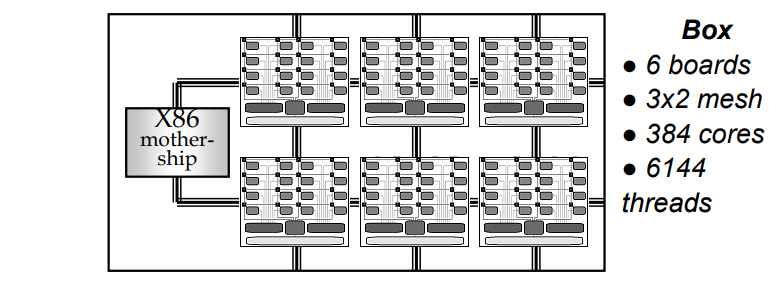}
\end{center}
\caption{POETS Box Layout}
\label{fig:hierarchy3}
\end{figure}

\begin{figure}[h!]
\begin{center}
	\includegraphics[scale=0.72]{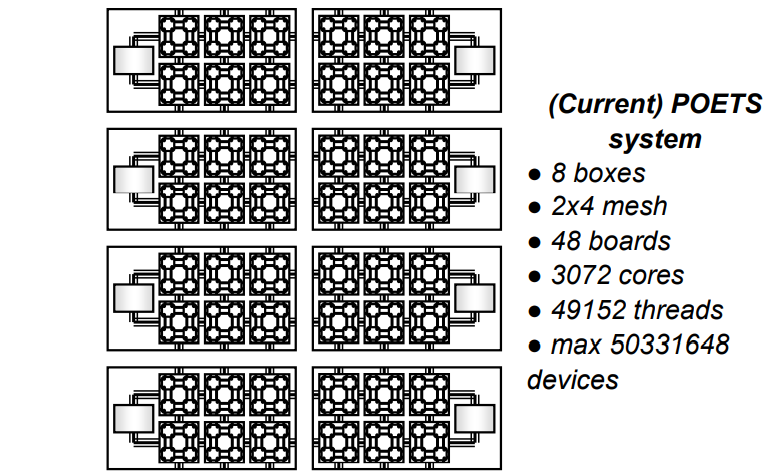}
\end{center}
\caption{Current POETS Cluster Layout}
\label{fig:hierarchy4}
\end{figure}

\indent POETS is a platform centered around a network-optimised FPGA cluster containing thousands of custom multi-threaded RISC-V cores [8], arranged in a hierarchy with three major tiers consisting of the physical hardware and two software APIs. \\
\indent An overview of the system is shown in Figures \ref{fig:hierarchy1}-\ref{fig:hierarchy4}. Beginning with the lowest level in the hardware tier, a basic hardware unit consists of four cores sharing a mailbox, cache and floating-point unit (FPU). This is called a tile and is depicted in Figure \ref{fig:hierarchy1}. Each core is instantiated as a customized 32-bit multi-threaded processor implementing a subset of the RV32IMF profile of the RISC-V instruction set, complete with 16 hardware threads. Each hardware thread is capable of scheduling multiple software threads, which are typically used to represent a single vertex in a graph-based solution. \\
\indent Each FPGA board (Stratix-V DE5-net) consists of 16 networked hardware tiles arranged in a 4 X 4 matrix sharing 4GB of off-chip RAM. This is shown in Figure \ref{fig:hierarchy2}. Four 10Gbps links are available for inter-board routing allowing an arbitrary number of boards to be networked. \\ 
\indent Six boards are placed into a box in a 3 X 2 matrix. This configuration was chosen due to thermal considerations and is depicted in Figure \ref{fig:hierarchy3}. External access for graph configuration and data entry/retrieval is provided via an x86 machine housed in the same box. High speed 10Gbps Ethernet links provide inter-box communication. \\
\indent Boxes may then be added and arranged arbitrarily, allowing for additional resources to fit the expansive requirements of large graph-based problem sets. Each box allows for 6144 hardware threads. The current POETS cluster consists of 8 boxes arranged in a 2 X 4 configuration, a total of 48 FPGAs allowing for 49,152 truly parallel hardware threads. This is shown in Figure \ref{fig:hierarchy4}.\\
\indent The lowest level in the software stack is a custom overlay called Tinsel\cite{NAYLOR}. This handles resource sharing, low level event handling and communication. Messages are routed via an unordered, guaranteed delivery approach. Send requests are arbitrated by the POETS infrastructure based on whether the network can guarantee delivery. Should there be insufficient network capacity at the moment of the send request, the sender can wait until there is sufficient resources to guarantee the send, temporarily store the message and continue alternative tasks until the network has capacity to guarantee the send or permanently abandon the send attempt. One-to-many event routing is enhanced by Tinsel’s distributed hardware multicast functionality\cite{NAYLOR3}. Tinsel also implements termination detection\cite{NAYLOR2} driven by vertices indicating when they have no more messages to send. This feature can be used to time-step globally synchronous applications as well as indicate the end of a processing run. \\
\indent On top of the Tinsel overlay sits a user-oriented API called POLite. POLite is a simple framework that allows for the design of vertices and inter-node communication channels (edges). The model is similar to that of Google’s Pregel\cite{MALEWICZ}. Handlers are defined for message sending/reception as well as cluster initialization, synchronized time-stepping and process termination.

\subsection{Configuration}

The current configuration of the POETS cluster (as described in the previous section) arranges the underlying FPGA fabric in a large, 2D array of RISC-V cores. Mapping of arbitrary graphs to this topology may be conducted manually or automatically. 
Applications for POETS may be written in either Tinsel or POLite. Tinsel uses fixed hardware addressing of the mailboxes combined with identifiers for governed threads and graphs may only be mapped manually. The application graph required to solve genotype imputation in a distributed manner is also a 2D array. This makes manually mapping the graph to the hardware threads relatively straightforward. POLite uses a definition of the graph and automatically maps vertices to threads using the METIS algorithm. In both cases the vertices and edges are stored in the board DRAM described in the previous section. Once the entire graph has been mapped, all cores with occupied threads are initialised and a ‘start’ event is globally broadcast to initiate execution.

\section{Event-Driven Implementation}

\subsection{Application Graph}

\indent Existing implementations of the Li and Stephens model are x86 based. The reference panels are stored in RAM which creates a large memory overhead (10’s of GB per thread). A single x86 thread is used to impute the markers for a single target haplotype by iteratively performing the calculations of the algorithm over the reference panel. \\
\indent In a distributed architecture like POETS, this idea has to be completely reworked. The reference panel is distributed across the cluster, with each state given its own vertex in the 2D graph. Each vertex is provided with the reference base, haplotype and marker number from the reference panel, as well as the genetic distance, $d_{m}$, to the markers contained in the previous column. They are then loaded with the annotated bases from the target haplotype, if one exists, at that marker location. \\

\subsection{Event-Driven Algorithm}

All calculations are performed in an event-driven manner by the vertices and message passing. The proposed event-driven algorithm can be seen in Algorithm 1.

 \begin{algorithm}[h!]
 \caption{Event-Driven Imputation}
 \begin{algorithmic}[1]
 \renewcommand{\algorithmicrequire}{\textbf{Input:}}
 \renewcommand{\algorithmicensure}{\textbf{Output:}}
 \REQUIRE msgType, h, match, $\alpha/\beta$ 
 \ENSURE  msgType, h, match, $\alpha/\beta$
 \\ \textit{\textbf{Initialization}} :
  \STATE Calculate $\tau_{m}$, same/diff $a_{ij}$. Inject first target haplotype.
  \STATE \textbf{if} (m = 1) $\alpha \leftarrow 1/|H|$; \textbf{if} (m = M) $\beta \leftarrow 1$;
  \STATE \textbf{Send request} to multicast known $\alpha/\beta$ forward/backward.
 \\ \textit{\textbf{Event Triggered Handlers}} :
 \\ \textit{\textbf{Received Message}} :
  \IF {($msgType = alpha$)}
  \STATE Accumulate $\alpha$ * $a_{ij}$ ($a_{ij}$ depends on h)
  \IF {($msgCnt = |H|$)}
  \STATE $\alpha \leftarrow \alpha * b_j(O_{m+1})$.
  \STATE \textbf{Send request} to multicast calculated $\alpha$ forward.
  \IF {($\alpha$ and $\beta$ known)}
  \STATE Calculate Posterior. \textbf{if} (h $\ne$ H) \textbf{Send request} to unicast Posterior downward
  \ENDIF
  \ENDIF
  \ENDIF
  \IF {($msgType = beta$)}
  \STATE Accumulate $a_{ij}$ * $\beta$ * $b_j(O_{m+1})$ ($a_{ij}$ depends on h) ($b_j(O_{m+1})$ = match)
  \IF {($msgCnt = |H|$)}
  \STATE \textbf{Send request} to multicast calculated $\beta$ backward.
  \IF {($\alpha$ and $\beta$ known)}
  \STATE Calculate Posterior. \textbf{if} (h $\ne$ H) \textbf{Send request} to unicast Posterior downward
  \ENDIF
  \ENDIF
  \ENDIF
  \IF {($msgType = posterior$)}
  \STATE Accumulate posterior based on allele label
  \ENDIF
  \\ \textit{\textbf{Step (No Active Send Requests)}} :
  \STATE Inject next target haplotype.
  \STATE \textbf{if} (m = 1) $\alpha \leftarrow 1/|H|$; \textbf{if} (m = M) $\beta \leftarrow 1$;
  \STATE \textbf{Send request} to multicast known $\alpha/\beta$ forward/backward.
 \end{algorithmic} 
 \end{algorithm}

\begin{figure}[h!]
\begin{center}
	\includegraphics[scale=0.51]{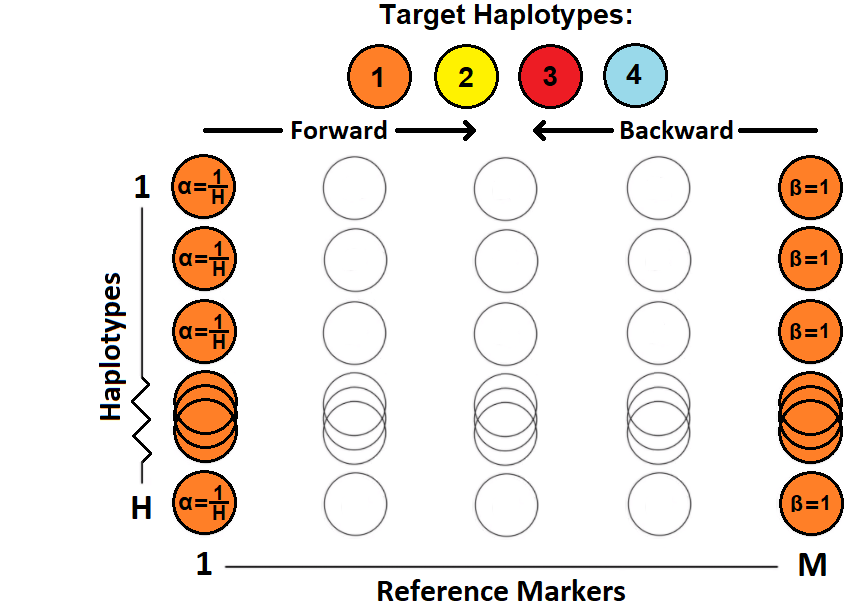}
\end{center}
\caption{Step 1 - Initial State}
\label{fig:step1}
\end{figure}

\begin{figure}[h!]
\begin{center}
	\includegraphics[scale=0.51]{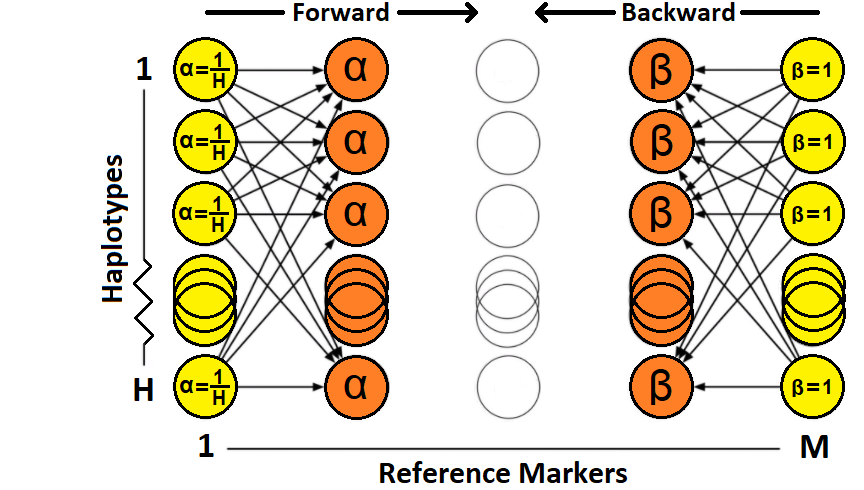}
\end{center}
\caption{Step 2 - First Alpha/Beta Calculation}
\label{fig:step2}
\end{figure}

\begin{figure}[h!]
\begin{center}
	\includegraphics[scale=0.51]{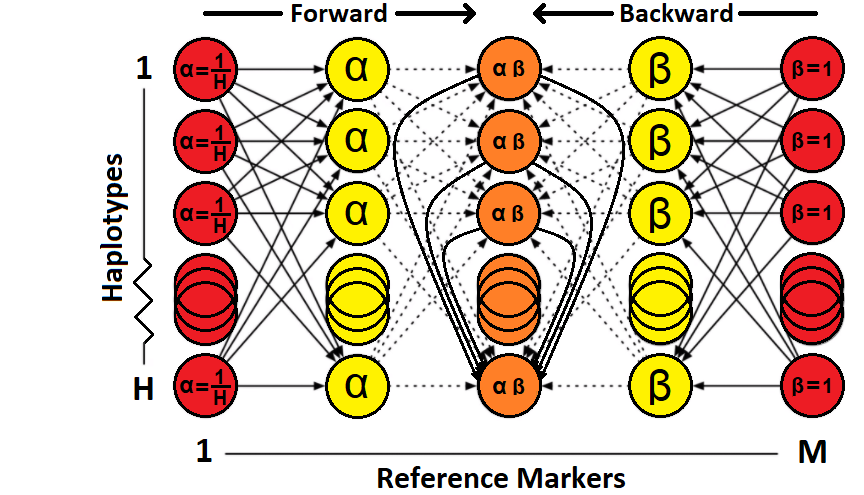}
\end{center}
\caption{Step 3 - First Posterior Probability Calculation}
\label{fig:step3}
\end{figure}

\begin{figure}[h!]
\begin{center}
	\includegraphics[scale=0.51]{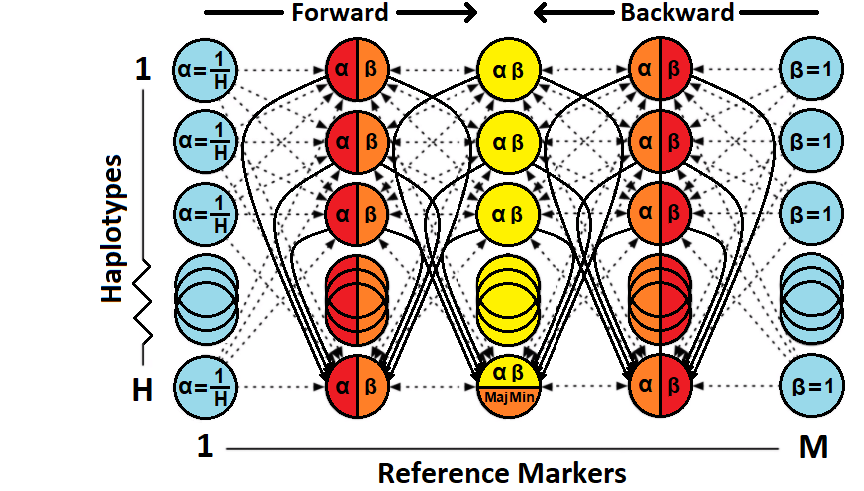}
\end{center}
\caption{Step 4 - First Allele Identification}
\label{fig:step4}
\end{figure}

\indent A step-by-step illustration of the algorithm in progress may be seen in Figures \ref{fig:step1}-\ref{fig:step4}. In the first step (Figure \ref{fig:step1}), all alpha values in the first marker location (column 1) are initialized to $1/|H|$ and all betas in the final marker location (column M) are initialized to 1 as described in Section \ref{section:model}. These values correspond to the first target haplotype. The vertices generate send requests for their respective alpha/beta value using the accelerated multicast functionality. \\
\indent In step two (Figure \ref{fig:step2}), the send requests generated in the previous step are serviced whilst there is network capacity. Vertices in the second marker location (column 2) receive the alpha values from the messages ($\alpha_{m}(i)$), and multiply them by the appropriate transition probability, $\alpha_{ij}$, which is determined by the genetic distance between the two marker locations, $d_{m}$, and whether both the sending and receiving vertex are on the same haplotype via the Tau factor. Once all alpha values from the messages have been accumulated, each vertex applies the appropriate emission probability, $b_{j} (O_{m+1})$, if a marker in the target haplotype is annotated at that marker location. The alpha values for vertices in the second marker location have now been calculated. In parallel, the alpha values in the first marker location (column 1) are initialized to $1/|H|$. These correspond to the alpha values for the second target haplotype. \\
\indent Simultaneously, vertices in the penultimate marker location (column M-1) receive the beta values from the messages ($\beta_{m}(i)$), and multiply them by the appropriate emission probability, $b_{j} (O_{m+1})$, and appropriate transition probability, $\alpha_{ij}$, in a similar manner to the alpha values just discussed. Once accumulated, the beta values for the vertices in the penultimate marker location have now been calculated. In parallel, the beta values in the final marker location (column M) are initialized to 1. These correspond to the beta values for the second target haplotype. The vertices in all four marker locations (first, second, penultimate, final) generate send requests for their respective alpha/beta value using the accelerated multicast functionality. \\
\indent In step three (Figure \ref{fig:step3}), all send requests generated in the previous step are serviced whilst there is network capacity. Vertices in the second and third marker locations (columns 2,3) receive the alpha values from the messages and calculate their own alpha values as described in step two. Vertices in the third and penultimate marker locations (columns 3,4) receive the beta values from the messages and calculate their own beta values as described in step two. Vertices in the third marker location now have both an alpha and beta value corresponding to the first target haplotype. These are multiplied together to generate a posterior probability for that vertex. A send request is generated to send this value, along with the allele labelled at that vertex, to the vertex in the final haplotype (H) at that marker location (column) as a unicast message. The alpha/beta value in the first/final marker locations are initialised as described in step one. These correspond to values for the third target haplotype. Vertices in all five marker locations generate the necessary send requests to propagate their known alpha/beta values using the accelerated multicast functionality. \\
\indent In step four (Figure \ref{fig:step4}), the send requests generated in the previous step are serviced whilst there is network capacity. The vertex in the final haplotype (H) of the third marker location (column 3) receives the unicast messages and accumulates the posterior probabilities based on the allele labels. Once all of the unicast messages have been received, the vertex knows which allele has the highest accumulated probability and therefore knows which allele should reside at that marker location for the first target haplotype (these are labelled major/minor in diallelic data). All vertices at marker location three also receive multicast messages to calculate the alpha/beta values corresponding to target haplotype two as described in step three. All vertices in the second and penultimate marker location (columns 2/4) know both the alpha/beta values for target haplotype one from step two and the alpha/beta calculated in this step. They therefore calculate the posterior probabilities for those marker locations and generate unicast send requests as described in step three. The alpha/beta value in the first/final marker locations are initialised as described in step one. These correspond to values for the fourth target haplotype. Vertices in all five marker locations generate the necessary send requests to propagate their known alpha/beta values using the accelerated multicast functionality.

\indent The algorithm proceeds until the accumulated allele probabilities have been calculated for all target haplotypes. All multicast messages contain both the alpha/beta value and reference haplotype number. The appropriate transition probability, $\alpha_{ij}$, is then applied by the receiving vertex. Doing this allows the alpha/beta to be the same for all receiving vertices and therefore enables the use of the accelerated multicast. As each vertex needs to receive a value from all vertices in the previous column, there is a natural synchronicity to the application. Experimentation demonstrated that using the POETS termination detection to synchronize the steps increases the average timestep by only 3\%, a more favourable penalty than adding additional logic to prevent cross contamination from multiple target haplotypes in a purely asynchronous implementation.

\subsection{Linear Interpolation}

\begin{figure}[h!]
\begin{center}
	\includegraphics[scale=0.585]{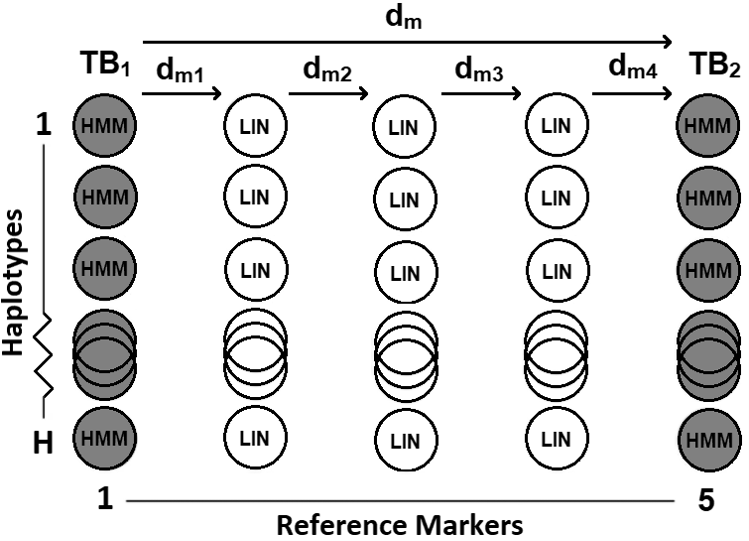}
\end{center}
\caption{Linear Interpolation}
\label{fig:linear}
\end{figure}

There are several optimisations that have been explored in the literature designed at reducing the computational complexity of genomic imputation. One optimisation with a significant impact has proven to be linear interpolation\cite{BROWNING2}. This takes advantage of the emission probability, $b_{j} (O_{m+1})$, falling from the alpha/beta equations where no base is annotated from the target haplotype. By only performing the Hidden Markov Modelling on marker location with annotated bases from the target haplotype, linear interpolation may be used to calculate the states in between. \\
\indent This concept is demonstrated in Figure \ref{fig:linear}. Reference markers 1 and 5 have annotated markers from the target haplotype ($TB_{1}$, $TB_{2}$). The Li and Stephens model is performed from marker (column) 1 to marker (column) 5 using the accumulated genetic distance between them, $d_{m}$. Once the values of these locations are known, the intermediate marker locations (2,3,4) may be estimated by apportioning a fraction of the accumulated change in accordance with the proportionality of the component genetic distances that make up $d_{m}$. \\
\indent This methodology is considerably less computationally complex than performing HMM on all states and has been consistently shown to deliver significant performance improvement in exchange for a negligible impact on the accuracy of the results (for genuine imputation upscale factors which can range from $\mathtt{\sim}$2 to $\mathtt{\sim}$170, HapMap3\cite{Altshuler} sampled $\mathtt{\sim}$1.4M markers, TopMED\cite{TOPMED} sampled $\mathtt{\sim}$240M).

\section{Experiments}

\subsection{x86 Baseline Implementation}

There are several x86 implementations of genotype imputation that have been explored in the literature. Many have been in development for well over a decade and therefore include many optimisations for the x86 platform. These include windowing, checkpointing, linear interpolation\cite{BROWNING2}, haplotype clustering\cite{BROWNING3, BROWNING4}, identity-by-descent\cite{LIU, ZHOU} and a host of other techniques that do not necessarily transfer into the distributed domain. However, the most computationally complex component still remains the Hidden Markov modelling. To provide the fairest comparison, a baseline x86 genotype imputation application was also written in C. This allowed multiple experiments to be conducted with optimisation levels that match in both the x86/event-driven versions. The C version contains three simple for loops. The first calculates an alpha/beta value from the relevant values. This is then nested inside a second loop that iterates over each haplotype (row) in the reference panel. This code is then nested inside a third loop that iterates over each marker (column) in the reference panel. The implementation first calculates the alphas then the betas.

\begin{figure}[h!]
\begin{center}
	\includegraphics[width=8.5cm,height=7.2cm]{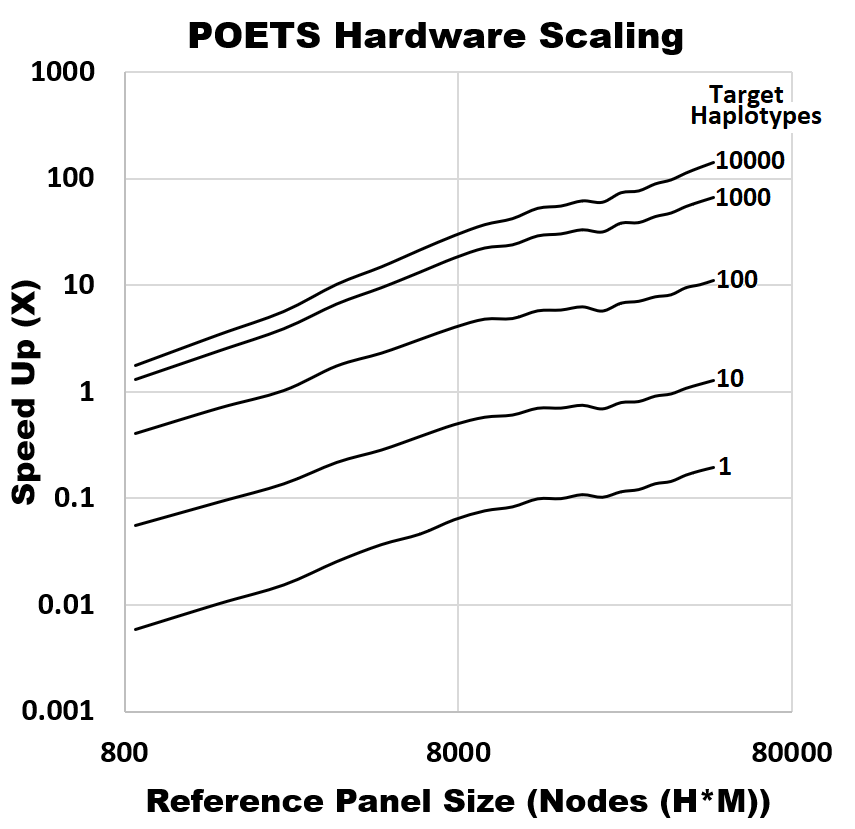}
\end{center}
\caption{Event-driven Algorithm Over Expanding Hardware}
\label{fig:expresults}
\end{figure}

\begin{figure}[h!]
\begin{center}
	\includegraphics[width=8.5cm,height=7.2cm]{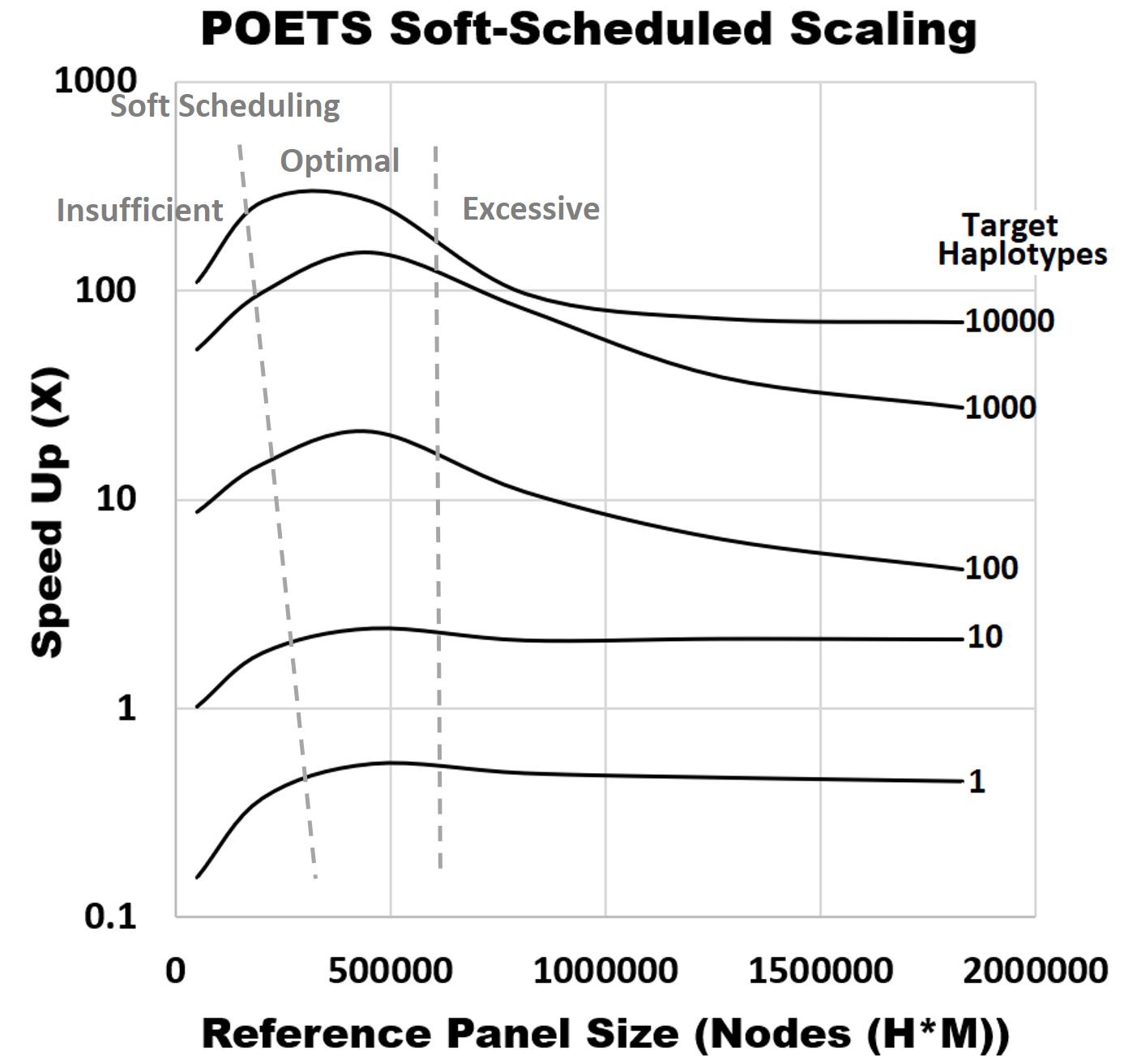}
\end{center}
\caption{Event-driven Algorithm Over Increased Soft-Scheduling}
\label{fig:softresults}
\end{figure}

It then calculates the posterior probabilities and accumulates them to determine the allele frequencies. This was run on an Intel i9-7940X CPU @ 3.10GHz. By comparison, the RISC-V cores of the POETS cluster are clocked at only 210MHz.

\subsection{Raw Model}

Two experiments were conducted to test the performance of the proposed event-driven algorithm. The first was designed to evaluate how the new event-driven algorithm scales into expanding hardware resources. Reference panel sizes less than the 49,152 hardware threads available from all 48 FPGAs were generated using features from genuine GWAS. Genetic distances were generated using a randomized uniform distribution seeded from HapMap3 data. Diallelic data was generated with an overall minor allele frequency of 5\%, widely regarded as the cut off for genotype estimation. A legitimate target marker to reference marker ratio of 1/100 was used. The aspect ratio of the reference panels was chosen based on haplotypes/markers
in existing GWAS, assuming genotyping technology chooses markers for a uniform distribution and noting that chromosome 1 accounts for approximately 8\% of the whole human genome. Varying numbers of target haplotypes were batch processed, the POETS wall-clock time was recorded and this was compared to the x86 implementation running the same reference panels. The results may be seen in Figure \ref{fig:expresults}. These demonstrate a clear and consistent positive trend in performance increase comparative to the single-threaded x86 implementation as the reference panel sizes are increased. The suggestion therefore is that this trend would continue should more hardware resources (FPGAs) be added to the cluster.

\indent The second experiment was designed to evaluate how the new event-driven algorithm responds to soft-scheduling. In this case, multiple vertices were assigned to each hardware thread, in order to increase the reference panel sizes that the cluster could handle. The full cluster (48 FPGAs) was used and the smallest reference panel size was chosen to just exceed the number of free hardware threads (49,152). The reference panels were generated with the same assumptions as the previous experiment. The results may be seen in Figure \ref{fig:softresults}. These suggest an optimal region of soft-scheduling exists at approximately 10 reference panel states per hardware thread, with insufficient or excessive soft-scheduling resulting in a diminished comparative speed up. A speed up of 270X for 10000 target haplotypes peaks in this region. Interestingly, excessively soft-scheduling reference panel states to fit ever larger reference panels onto the cluster still demonstrates superior performance over the x86 implementation. This is important as the limiting factor in the real-life problem is handling the expansive reference panel sizes.

\begin{figure}[h!]
\begin{center}
	\includegraphics[scale=0.57]{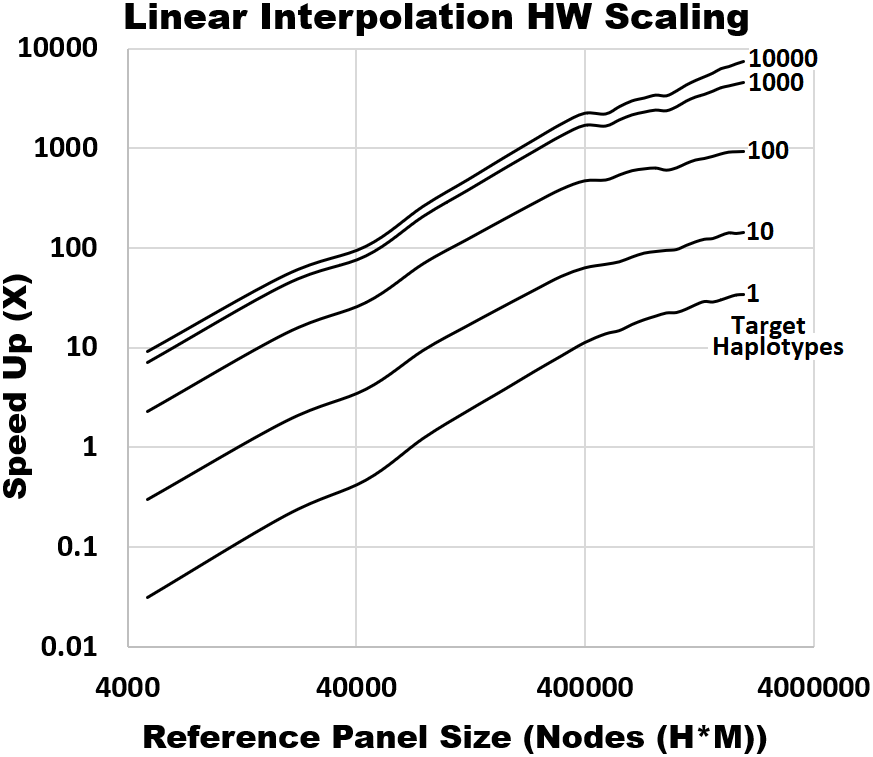}
\end{center}
\caption{Linear Interpolation Algorithm Over Expanding Hardware}
\label{fig:linresults}
\end{figure}

\subsection{Linear Interpolation}

Linear interpolation was then added into the distributed algorithm (and also the baseline x86 implementation). Reference panels were generated with the same assumptions used in the previous two experiments, with the exception of lowering the target marker to reference marker ratio to 1/10. This was done to accommodate the RAM allocation to each board as described in the POETS architecture. Reference panels of varying sizes were generated. In those with sizes less than 49,152, each hardware thread governed a state section consisting of a single HMM state and 9 linear interpolation states. In reference panels with sizes greater than 49,152, one/many/all hardware threads governed multiple HMM/Linear Interpolation state sections. The results may be seen in Figure \ref{fig:linresults}. The largest reference panel size that the cluster was capable of accepting was the same size as in the previous experiment. This is due to the fact that the limiting factor is the memory required to store the reference panel. As each vertex now governed 10 references panel states, the memory requirement per vertex has increased, but the number of messages required to solve the reference panel as a whole has decreased by a similar factor ($\mathtt{\sim}$10X). This is important as the queuing and handling of hundreds of messages per receiving vertex (the fan in) in the raw algorithm was likely the factor limiting performance on the POETS cluster.\\
\indent This HMM/Linear interpolation multi-tasking per vertex means that the number of vertices per hardware thread is also $\mathtt{\sim}$10X less than the previous experiment at the largest reference panel size (4 vs 40). As the distributed/x86 comparative wall-clock time consistently improves, this suggests that the optimal soft-scheduling region is not reached prior to memory exhaustion and that allowing for larger reference panels simply by increasing the physical memory in the current POETS architecture would demonstrate greater performance improvement up until the optimal soft-scheduled region. \\ 
\indent Based on the current event-driven implementation and manually accounting for the memory requirements in the Tinsel layer, genuine reference panels would require a POETS cluster $\mathtt{\sim}$16X larger than the current hardware. A next generation cluster with significantly improved hardware (based on Intel Stratix 10's) is currently under construction. This should include a ($\mathtt{\sim}$6.5X) increase in hardware thread count, a 2X increase in core frequency, an 8X increase in DRAM per board complete with a 2X increase in bandwidth per memory chip and a 10X increase in inter-board communication bandwidth. All of these factors should significantly enhance the performance of the event-driven implementation.

\section{Final Comments}
This work proposed and investigated an event-driven solution to genotype imputation by implementing a new event-driven algorithm of the widely accepted Li and Stephens model. This was run on a custom RISC-V NoC FPGA cluster called POETS and compared to an x86-based implementation. The raw algorithm’s wall-clock time reduction comparative to the x86 implementation consistently improves over expanding hardware resources (scalability was demonstrated). Results also demonstrated that this comparative performance increase may be enhanced by soft-scheduling multiple reference panel states per hardware thread, allowing for the processing of reference panel sizes beyond the optimal operating point. Reductions in wall--clock time of 270X were demonstrated. \\
\indent Linear interpolation optimisation was then introduced. Whilst this did not allow reference panels of a larger size to be solved (the limiting factor was the memory required to store the panel), the wall-clock times of those that were solvable were significantly improved due to the reduction in the number of messages required. This factor exceeded the upscale factor used (1/10) and resulted in comparative performance improvement of $\mathtt{\sim}$5 orders of magnitude when compared to a single-threaded x86 solution with similar optimisation. \\
\indent Overall, the work presented suggests that genotype imputation may warrant further investigation in the event-driven domain.

% if have a single appendix:
%\appendix[Proof of the Zonklar Equations]
% or
%\appendix  % for no appendix heading
% do not use \section anymore after \appendix, only \section*
% is possibly needed

% use appendices with more than one appendix
% then use \section to start each appendix
% you must declare a \section before using any
% \subsection or using \label (\appendices by itself
% starts a section numbered zero.)
%

% use section* for acknowledgment
\ifCLASSOPTIONcompsoc
  % The Computer Society usually uses the plural form
  \section*{Acknowledgments}
\else
  % regular IEEE prefers the singular form
  \section*{Acknowledgment}
\fi

This work was supported under EPSRC Grant 
EP/N031768/1 (POETS). The corresponding author is Jordan Morris.

% Can use something like this to put references on a page
% by themselves when using endfloat and the captionsoff option.
\ifCLASSOPTIONcaptionsoff
  \newpage
\fi

% trigger a \newpage just before the given reference
% number - used to balance the columns on the last page
% adjust value as needed - may need to be readjusted if
% the document is modified later
%\IEEEtriggeratref{8}
% The "triggered" command can be changed if desired:
%\IEEEtriggercmd{\enlargethispage{-5in}}

% references section

% can use a bibliography generated by BibTeX as a .bbl file
% BibTeX documentation can be easily obtained at:
% http://mirror.ctan.org/biblio/bibtex/contrib/doc/
% The IEEEtran BibTeX style support page is at:
% http://www.michaelshell.org/tex/ieeetran/bibtex/
%\bibliographystyle{IEEEtran}
% argument is your BibTeX string definitions and bibliography database(s)
%\bibliography{IEEEabrv,../bib/paper}

\bibliographystyle{unsrt}
\bibliography{bare_adv.bib}

% <OR> manually copy in the resultant .bbl file
% set second argument of \begin to the number of references
% (used to reserve space for the reference number labels box)
%\begin{thebibliography}{1}

%\bibitem{IEEEhowto:kopka}
%H.~Kopka and P.~W. Daly, \emph{A Guide to {\LaTeX}}, 3rd~ed.\hskip 1em plus
%  0.5em minus 0.4em\relax Harlow, England: Addison-Wesley, 1999.

%\end{thebibliography}

% biography section
% 
% If you have an EPS/PDF photo (graphicx package needed) extra braces are
% needed around the contents of the optional argument to biography to prevent
% the LaTeX parser from getting confused when it sees the complicated
% \includegraphics command within an optional argument. (You could create
% your own custom macro containing the \includegraphics command to make things
% simpler here.)
%\begin{IEEEbiography}[{\includegraphics[width=1in,height=1.25in,clip,keepaspectratio]{mshell}}]{Michael Shell}
% or if you just want to reserve a space for a photo:

%\clearpage
\vspace{-1cm}

\begin{IEEEbiography}[{{\includegraphics[clip,width=1in,height=1.25in,keepaspectratio]{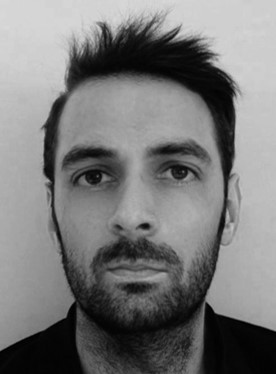}}}]{Jordan Morris}
(MEng `13, PhD `18 -- Newcastle University, UK) was previously a research fellow
at Arm as part of the Applied Silicon research group.
At present, he works as a Research Associate in the School of Engineering,
Newcastle University, with research interests in custom hardware and many-core systems focusing on machine learning and bioinformatic algorithms.
\end{IEEEbiography}

\vspace{-1.4cm}

\begin{IEEEbiography}[{{\includegraphics[clip,width=1in,height=1.25in,keepaspectratio]{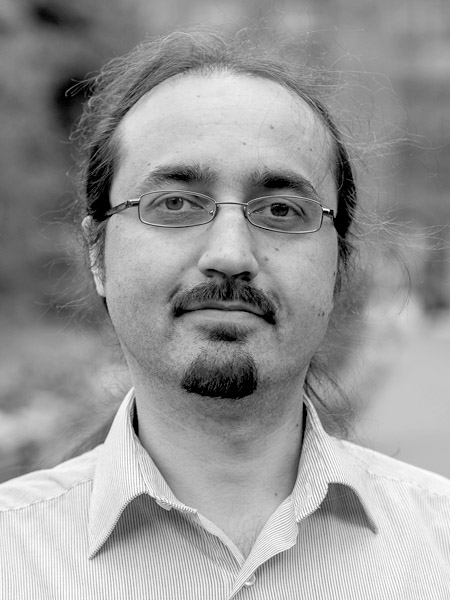}}}]{Ashur Rafiev}
(MEng `05, PhD `11 -- Newcastle University, UK) previously worked
in the Human-Computer Interaction and the Security and Software Reliability
research groups in the School of Computing, Newcastle University.
At the moment, he works as a Research Associate in the School of Engineering,
Newcastle University, with the research interest in many-core systems
including power modeling, simulation, and applications for distributed
computing.
\end{IEEEbiography}

\vspace{-1.4cm}

\begin{IEEEbiography}[{{\includegraphics[clip,width=1in,height=1.25in,keepaspectratio]{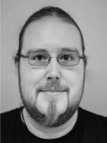}}}]{Graeme M. Bragg}
(MEng `12, PhD `17) has previously worked for Esri UK and is currently
employed by the University of Southampton as a Research Fellow on
the POETS project. He has 3 journal and 8 conference publications.
His research interests include environmental sensor networks, low-power
networks and many-core systems.
\end{IEEEbiography}

\vspace{-1.4cm}

\begin{IEEEbiography}[{{\includegraphics[clip,width=1in,height=1.25in,keepaspectratio]{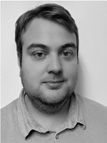}}}]{Mark L. Vousden}
(MEng `12, PhD `17) has been employed at BluPoint Ltd and is currently
a Research Fellow at the University of Southampton, UK. He has 6 journal
publications in micromagnetic simulation. Current research interests
are event-driven high-performance computing architectures and paradigms.
\end{IEEEbiography}

\vspace{-1.4cm}

\begin{IEEEbiography}[{{\includegraphics[clip,width=1in,height=1.25in,keepaspectratio]{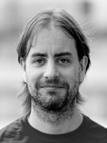}}}]{David B. Thomas}
(MSc `00, PhD `05, MEd `18) is a Senior Lecturer at Imperial College
London, working at the boundaries of software, processors, and custom
digital hardware. Interests include reconfigurable hardware, high-level
languages and libraries for FPGAs, optimised numerical algorithms
and IP cores.
\end{IEEEbiography}

\vspace{-1.4cm}

\begin{IEEEbiography}[{{\includegraphics[clip,width=1in,height=1.25in,keepaspectratio]{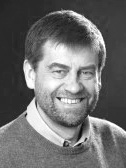}}}]{Alex Yakovlev}
is a professor in the School of Engineering, Newcastle University.
His research interests include asynchronous circuits and systems,
concurrency models, energy-modulated computing. Yakovlev received
a DSc in Engineering at Newcastle University. He is FIEEE, FIET, and
Fellow of RAEng. In 2011-2013 he was a Dream Fellow of the UK Engineering
and Physical Sciences Research Council (EPSRC). 
\end{IEEEbiography}

%\vspace{-3.68 cm}

% You can push biographies down or up by placing
% a \vfill before or after them. The appropriate
% use of \vfill depends on what kind of text is
% on the last page and whether or not the columns
% are being equalized.

%\vfill

% Can be used to pull up biographies so that the bottom of the last one
% is flush with the other column.
%\enlargethispage{2cm}

\begin{IEEEbiography}[{{\includegraphics[clip,width=1in,height=1.25in,keepaspectratio]{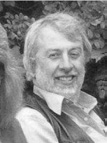}}}]{Andrew D. Brown}
(BSc `76, PhD `81) has held posts at IBM Hursley Park (UK), Siemens
NeuPerlach (Germany), Multiple Access Communications (UK), LME Design
Automation (UK), Trondheim University (Norway), Cambridge University
(UK), and EPFL (CH). He is a professor of electronics at Southampton
University, UK. He is FIET, FBCS, CEng, CITP, Eur Ing and SMIEEE.
\end{IEEEbiography}

% that's all folks
\end{document}